\documentclass[aps,prl,preprint,showpacs,superscriptaddress]{revtex4}
\usepackage{graphicx,epsfig}% Include figure files
\usepackage{amsmath}
\usepackage{amsfonts}
\usepackage{amssymb}
\usepackage{bm}% bold math
\usepackage{booktabs}

\newcommand{\rem}[1]{}

\begin{document}

\title{Complex magnetism of lanthanide intermetallics unravelled}
\author{L. Petit}
 \affiliation{ Daresbury Laboratory, Daresbury, Warrington WA4 4AD, UK }
\author{D. Paudyal}
 \affiliation{ The Ames Laboratory, US Department of Energy, Iowa State University, Ames, Iowa 50011-3020, USA }
\author{Y. Mudryk}
 \affiliation{ The Ames Laboratory, US Department of Energy, Iowa State University, Ames, Iowa 50011-3020, USA }
\author{K. A. Gschneidner Jr.}
\affiliation{ The Ames Laboratory, US Department of Energy and Department of Materials Science and
Engineering, Iowa State University, Ames, Iowa 50011-3020, USA }
\author{V. K. Pecharsky}
\affiliation{ The Ames Laboratory, US Department of Energy and Department of Materials Science and
Engineering, Iowa State University, Ames, Iowa 50011-3020, USA }
\author{M. L\"{u}ders}
\affiliation{ Daresbury Laboratory, Daresbury, Warrington WA4 4AD, UK }
\author{Z. Szotek}
\affiliation{ Daresbury Laboratory, Daresbury, Warrington WA4 4AD, UK }
\author{R. Banerjee}
\affiliation{ Department of Physics, University of Warwick, Coventry CV4 7AL, U.K. }
\author{J. B. Staunton}
 \affiliation{ Department of Physics, University of Warwick, Coventry CV4 7AL, U.K. }
 
%\date{\today}
\begin{abstract}
We explain a profound complexity of magnetic interactions of some technologically 
relevant gadolinium intermetallics using an ab-initio electronic 
structure theory which includes disordered local moments and strong 
$f$-electron correlations. The theory correctly finds GdZn and GdCd to be simple ferromagnets and predicts a remarkably 
large increase of Curie temperature with pressure of +1.5 K kbar$^{-1}$ for GdCd confirmed by our experimental measurements 
of  +1.6 K kbar$^{-1}$.   Moreover we find the origin of a ferromagnetic-antiferromagnetic competition in GdMg manifested 
by non-collinear, canted magnetic order at low temperatures. Replacing 35\% of  the Mg atoms with Zn removes this transition 
in excellent agreement with longstanding experimental data.
\end{abstract}
\maketitle

Lanthanide compounds play an increasingly important role in the development of novel 
materials for high-tech applications which range from mobile phones and radiation 
detectors to air conditioning and renewable energies. Much of this stems from their 
magnetic properties, so that they are  indispensable components in permanent 
magnets,~\cite{Burzo} magneto-responsive devices for solid state 
cooling~\cite{Gschneidner_RepProg} and other applications.
Common to all the lanthanide elements is their valence electronic 
structure which makes them chemically similar and also causes magnetic order. 
Lanthanide atoms are predominantly divalent (5d$^0$6s$^2$ valence electron configuration), becoming mostly trivalent
in a solid, donating three valence electrons to the 
electron glue in which the atomically-localised $f$-electron magnetic moments 
sit. The interaction between these moments derives from how the electron glue is 
spin-polarised. The longstanding RKKY~\cite{Kasuya} (Ruderman-Kittel-Kasuya-Yosida) free 
electron model of this electronic structure is typically used to try to explain the many 
features of the indirect coupling of the 4$f$-electron moments despite its rather poor 
representation of the narrow band 5$d$-states. The possible importance of the latter 
has already been inferred from some earlier electronic structure studies~\cite{Campbell,Postnikov,Buschow-Svechkarev,Khmel}. 

Whilst theoretical aspects of lanthanide magnetism are well understood at the 
phenomenological level, predictive first principles calculations are challenging 
owing to the complexities of the strongly correlated $f$-electrons and itinerant valence 
electrons along with the magnetic fluctuations generated at finite temperatures. In this letter we explore 
lanthanide compounds with an ab-initio theory based on Spin Density Functional Theory (SDFT) in which the self-interaction corrected 
(SIC) local spin density (LSD) method~\cite{PZ-SIC,Pederson} provides an 
adequate description of f-electron correlations~\cite{strange_nature,rex,fel} and the disordered 
local moment (DLM) theory~\cite{Gyorffy_DLM} handles the magnetic fluctuations.
We are able to give a quantitatively accurate description of the diverse magnetism of 
Cs-Cl (B2) ordered phases of Gd with Zn, Cd and Mg which we test against experimental data 
and show the complex role played by the spin-polarised valence electrons.  

Local moments of fixed magnitudes are assumed to persist to high temperatures and in lanthanide compounds are formed naturally 
from partially occupied localised 4f-electron states. The orientations of these moments fluctuate slowly compared to the dynamics
of the  valence electrons glue surrounding them.
By labelling these transverse modes by local spin polarisation axes
fixed to each lanthanide atom $i$, $\hat{\bf e}_i$, and using a
generalisation of SDFT~\cite{Gyorffy_DLM}(+SIC~\cite{LSIC,REnature}) for
prescribed orientational arrangements, $\{\hat{\bf e}_i\}$, we can determine
the {\it ab-initio}
energy for each configuration, $\Omega\{\hat{\bf e}_i\}$~\cite{JBS+BLG,RDLM-big,REnature,BLG-paper,SI} so that the configuration's probability 
at a temperature $T$ can be found. The magnetic state of the system is set by an average over all such configurations, 
appropriately weighted, and
specifies the magnetic order parameters, 
$\{ {\bf m}_{i} = \langle \hat{\bf e}_i \rangle \}$, where the magnitudes 
$m_i = |{\bf m}_{i}|$ range from $0$
for the high temperature paramagnetic (PM) (fully disordered) state to $1$ 
when the magnetic order is complete at $T=0$K.  A distribution where the order parameters are the same on every site, 
$\{{\bf m}_{i} = m_f\,\hat{z}\}$ say, describes
a ferromagnetically ordered (FM) state whereas one where the ${\bf m}_{i}$ alternate layer by layer between $m_a \hat{x}$ and 
$- m_a \hat{x}$ characterises an
antiferromagnetic (AF1) order. The free energy function
$\mathcal{F}(\{{\bf m}_{i}\})$,  written in terms of these magnetic order parameters, ${\bf m}_{i}$,  monitors magnetic phase transitions. It contains the effects of the
spin-polarised valence electronic structure which adapts to the type
and extent of magnetic order~\cite{CoMnSi,FeRh,SI}. 
For lanthanide materials DLM theory describes how valence electrons mediate the interactions
between the f-electron moments. These can turn out to be RKKY-like, but
can also show strong deviations from this picture as we find here for simple Gd-containing prototypes.

We start with GdZn, of particular interest in solid 
state cooling,~\cite{Vitalij} but also because we expect its electronic structure to be straightforward~\cite{Rusz_GdX,Buschow-Svechkarev}.
The Gd atoms occupy a simple cubic lattice of the CsCl(B2) ordered phase. Our first-principles SIC-LSD calculations find the ground state Gd-ion configuration to be
trivalent (Gd$^{3+}$), with seven localised f-states constituting a stable half-filled shell, in line with
Hund's Rules~\cite{strange_nature,rex,SI}. So Gd of all the heavy lanthanides has the relative simplicity of an S-state for 
its f-electrons, largely uncomplicated by crystal field effects and spin-orbit coupling. This permits
a clinical look at how the interactions between large 4f-magnetic moments are mediated by the valence electrons. These 
come from both the lanthanide (5d$^1$6s$^2$) and the 
post-transition metal Zn which has low-lying, nominally filled, d-shells 
(3d$^{10}$) added to its two s-electrons.
Our {\it ab-initio} DLM theory can thus investigate the effect of the lanthanide 
5d electrons hybridising weakly with 3d states. This touches on a very 
important aspect of many magnetic materials containing 
both rare earth and transition metal elements~\cite{Sanyal} where understanding the interplay between 
the localised lanthanide magnetic moments and the more itinerant 
magnetism originating from the transition metal d-electrons is paramount for the design 
of more efficient materials. 

Our DLM theory calculations for the paramagnetic state of GdZn produce
local moments of magnitude $\mu \approx$ 7.3$\mu_B$ on the Gd sites pointing in random directions so that there is no long range 
magnetic order, $\{{\bf m}_{i}=0\}$. The calculated paramagnetic susceptibility~\cite{REnature,JBS+BLG,SI}, $\chi({\bf q})$, 
with a maximum at wave-vector  ${\bf q}_{max} = (0,0,0)$,  shows that, in accord with experiment,
GdZn develops ferromagnetic (FM) order below a Curie temperature, $T_c$ = 184 K (at  theoretically determined 
lattice constant, $a=a_{th}=$ 6.62 a.u.), somewhat lower than the experimental value of 
T$_c$ = 270 K~\cite{Kanematsu_GdZn}(at $a=a_{exp}=$ 6.81 a.u.). 
We find that GdZn's T$_c$ gradually decreases under pressure, $P$, with 
calculated $\frac{dT_C}{dP}=$ -0.45 K kbar$^{-1}$ which agrees reasonably well with experimental 
value of -0.13 K kbar$^{-1}$ from the literature~\cite{Hiraoka} (Fig.1(a)). 
The negative $\frac{dT_C}{dP}$ is typical of many metallic magnets owing to pressure-induced 
band broadening and diminished energy benefit from spin polarising the valence 
electrons around the Fermi energy. 

Na\"{i}vely one might expect similar effects if Zn is replaced with isoelectronic Cd whose filled 4$d-$band states are simply 
more extended than the 3$d$'s of Zn. 
Our calculations, however, show something rather different. Whilst both theory and experiment find GdCd to be a simple 
ferromagnet like GdZn, with T$_c$ = 234 K ($a_{th}=$ 6.98 a.u.) and  265 K\cite{Alfieri_GdCd}($a_{exp}=$ 7.09 a.u.~\cite{SI}), in sharp contrast 
to its results for GdZn, theory predicts its T$_c$ 
to increase quite dramatically with pressure (Fig.1), i.e. a positive and rather large $\frac{dT_c}{dP}$. 
Owing to the paucity of reliable published experimental 
pressure data for GdCd,~\cite{GdCd-old-pressure-exp} we have carried out measurements~\cite{SI} to test this specific
prediction and a comparison between the calculated and experimentally observed T$_{c}$'s for GdCd  
as a function of pressure is shown in Fig.1(a). 
\begin{figure}[h]
\begin{center}
(a)\includegraphics[width=75mm,height=50mm]{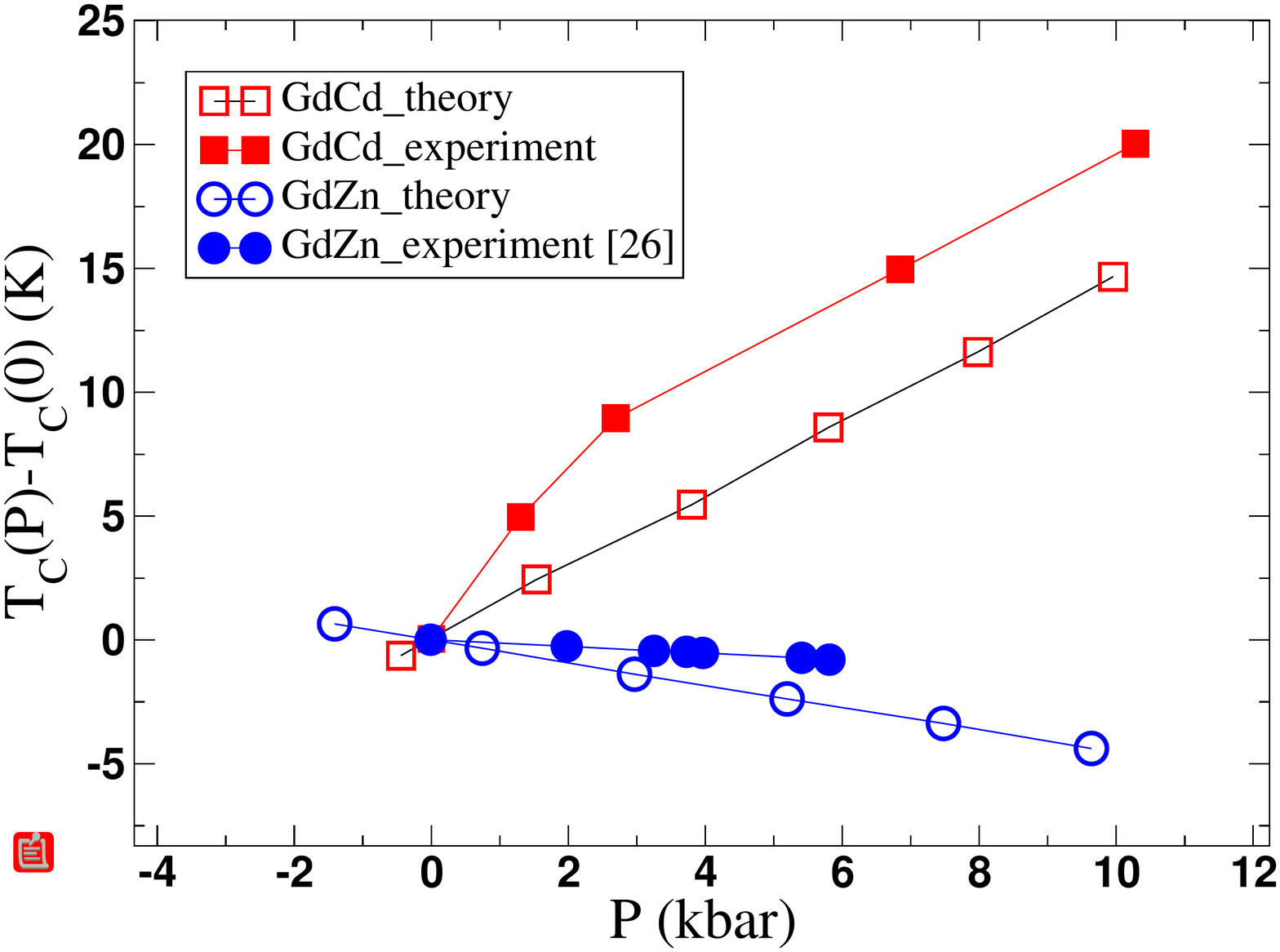}
(b)\includegraphics[width=80mm,height=50mm,clip]{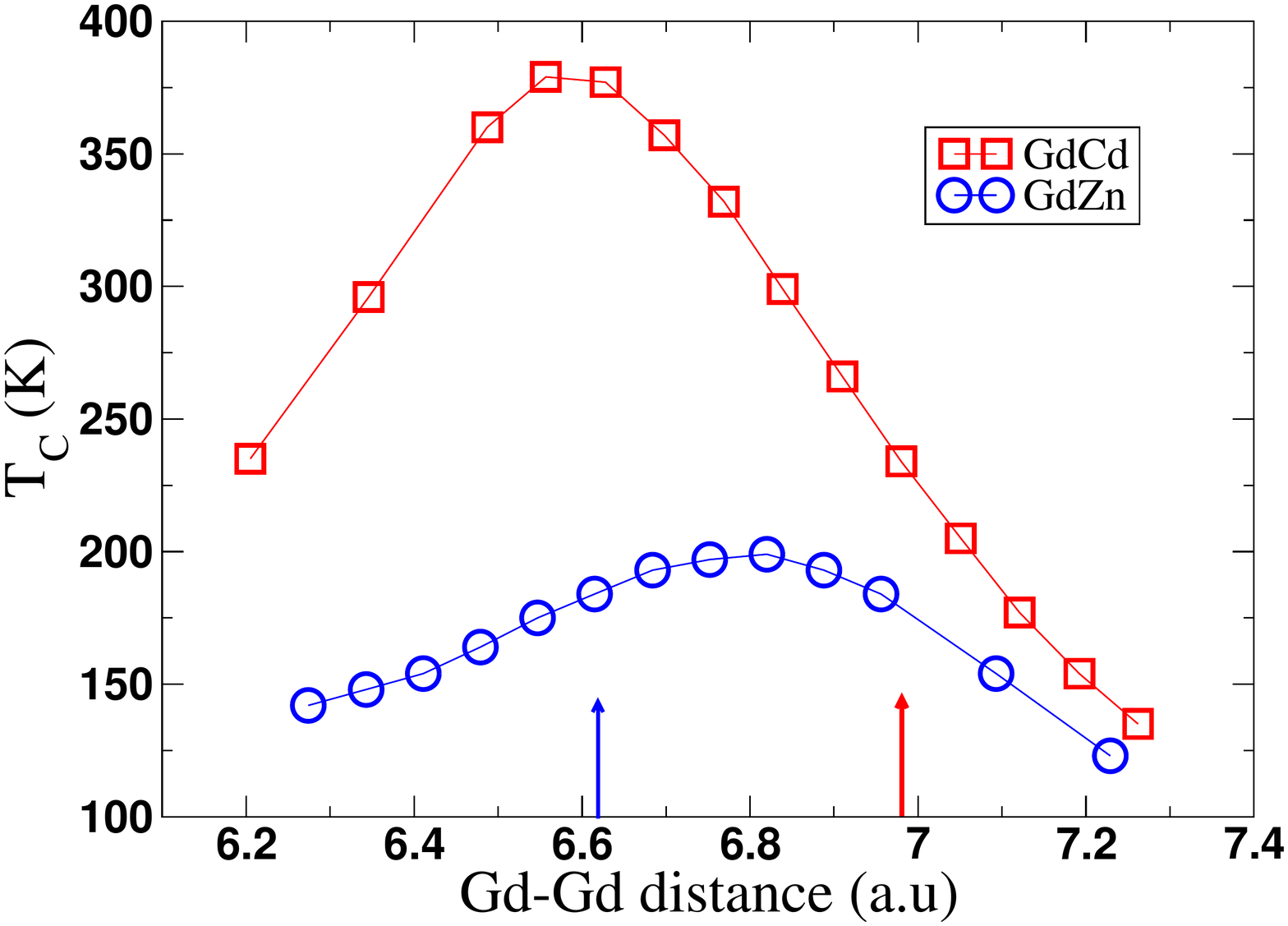}
\caption{(a) Comparison between theory (open symbols) and experimental~\cite{SI} (filled symbols) T$_c$ differences, 
$\left(T_c(P)-T_c(0)\right)$, as a function of pressure, $P$, 
for GdZn (blue circles) and GdCd (red squares). The experimental data for GdZn
are from Ref. (\onlinecite{Hiraoka}). (b) T$_c$ of GdZn (blue circles) and GdCd (red squares) as a function of lattice parameter $a$ (atomic units) calculated 
from the theory. The vertical arrows indicate $a_{th}$, red for GdCd and blue for GdZn.}
\label{pressure}
\end{center}
\end{figure}

The theory-experiment agreement is excellent:
$\frac{dT_c}{dP}$ from theory is +1.5 and from experiment is +1.6 K kbar$^{-1}$. 
Whilst not unusual for first-order magnetostructural transitions (e.g. $\approx$ 1-3 K kbar$^{-1}$ is 
observed in Gd$_5$Si$_x$Ge$_{4-x}$ alloys~\cite{Tseng}), this is a rather high rate 
for a second-order transition as occurs in GdCd. Reasons for this stark difference between GdZn and GdCd 
are found from our T$_c$ calculations as a function of lattice parameter, $a$,  
(Fig.1(b)). Starting from large values, T$_c$ initially increases with decreasing Gd-Gd 
distance for both GdZn and GdCd, reaching a maximum whence it starts decreasing with further reduction of the Gd-Gd distance.  
The $\frac{d T_c}{dP}$'s shown in Fig.1(a) originate from where the two 
compounds have their equilibrium lattice spacings, $a_{th}$, marked by blue (GdZn) and red 
(GdCd) arrows in Fig.1(b).   

For materials with the same number of valence electrons per atom, the RKKY account of magnetic interactions
would be the same.
GdMg is isoelectronic with both GdZn and GdCd but with no filled 3d or 4d band of states. This difference
 leads to our DLM theory finding GdMg's PM state to be 
qualitatively different than GdZn's and GdCd's.  
We find a discordant blend of FM and AF1 dominant magnetic correlations in the PM state 
- the calculated paramagnetic $\chi({\bf q})$ has two comparable peaks at wavevectors $(0,0,0)$ and  $(0,0,\frac{1}{2})$~\cite{SI}
(in units of $\frac{2 \pi}{a}$). 
Which one is stronger depends on the $a$ values used. 
At the theory volume ($a_{th}=$7.00 a.u. [c.f. $a_{exp}=$ 7.20 a.u.~\cite{GdMg-exp}]), our calculations predict a FM state below 
$T_c=$128K. 
Reducing the Gd-Gd separation weakens the FM aspects and, for example,  a  4$\%$ decrease
leads to an AF1 state instead, below the N\'{e}el temperature $T_N=$87K.

We determine the magnetic order that evolves as $T$ is lowered through the transition temperature to $0$K as a 
consequence of these competing FM and AF1 effects by using our DLM theory~\cite{SI} for the first time to describe
a magnetically ordered state with a canted structure and repeating the analysis for a number
of $a$ values. We set the order parameters, ${\bf m}_i$'s for 
the system at various stages of partial onto complete magnetic order, to alternate between $m_{f}\hat{z} + m_{a}\hat{x}$ and 
$m_{f}\hat{z} - m_{a} \hat{x}$ on consecutive Gd layers along the $(1,0,0)$ direction giving a canting angle between layers of
$\Theta_c= 2\arctan (\frac{m_a}{m_f})$ so that the overall magnetization of a system is local moment size $\mu$ (7.3 $\mu_B$)
times $m_f$. $m_f \neq 0$, $m_a =0$, $\Theta_c=0$  
signifies a FM state and $m_a \neq 0$, $m_f =0$, $\Theta_c=$180$^{\circ}$ a AF1 state. 
\begin{figure}[h]
\begin{flushleft}
(a) \includegraphics[trim=10mm 0 0 0,clip,width=90mm,height=45mm]{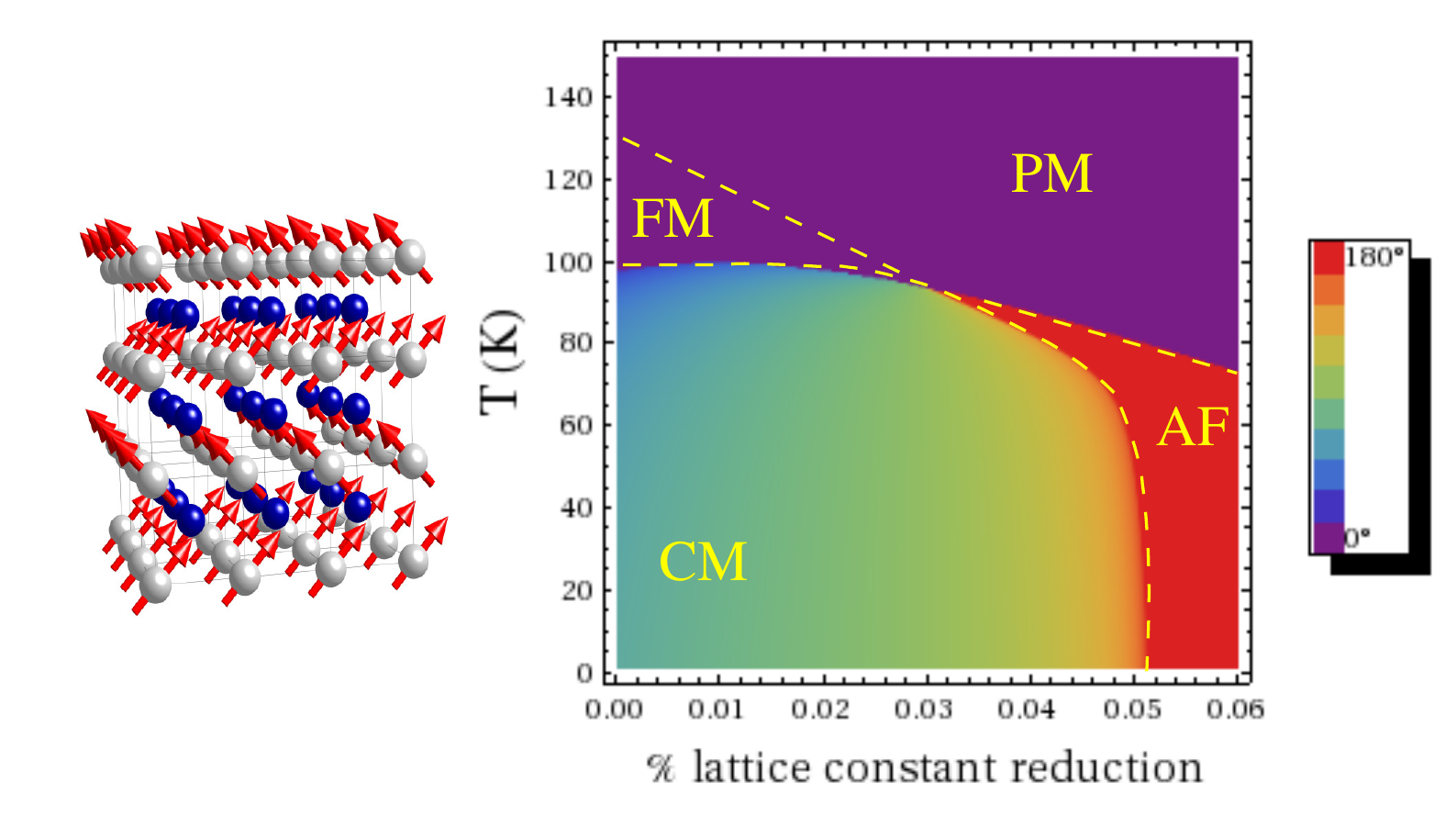}
(b)\includegraphics[trim=0 0 0 0,clip,width=65mm,height=45mm]{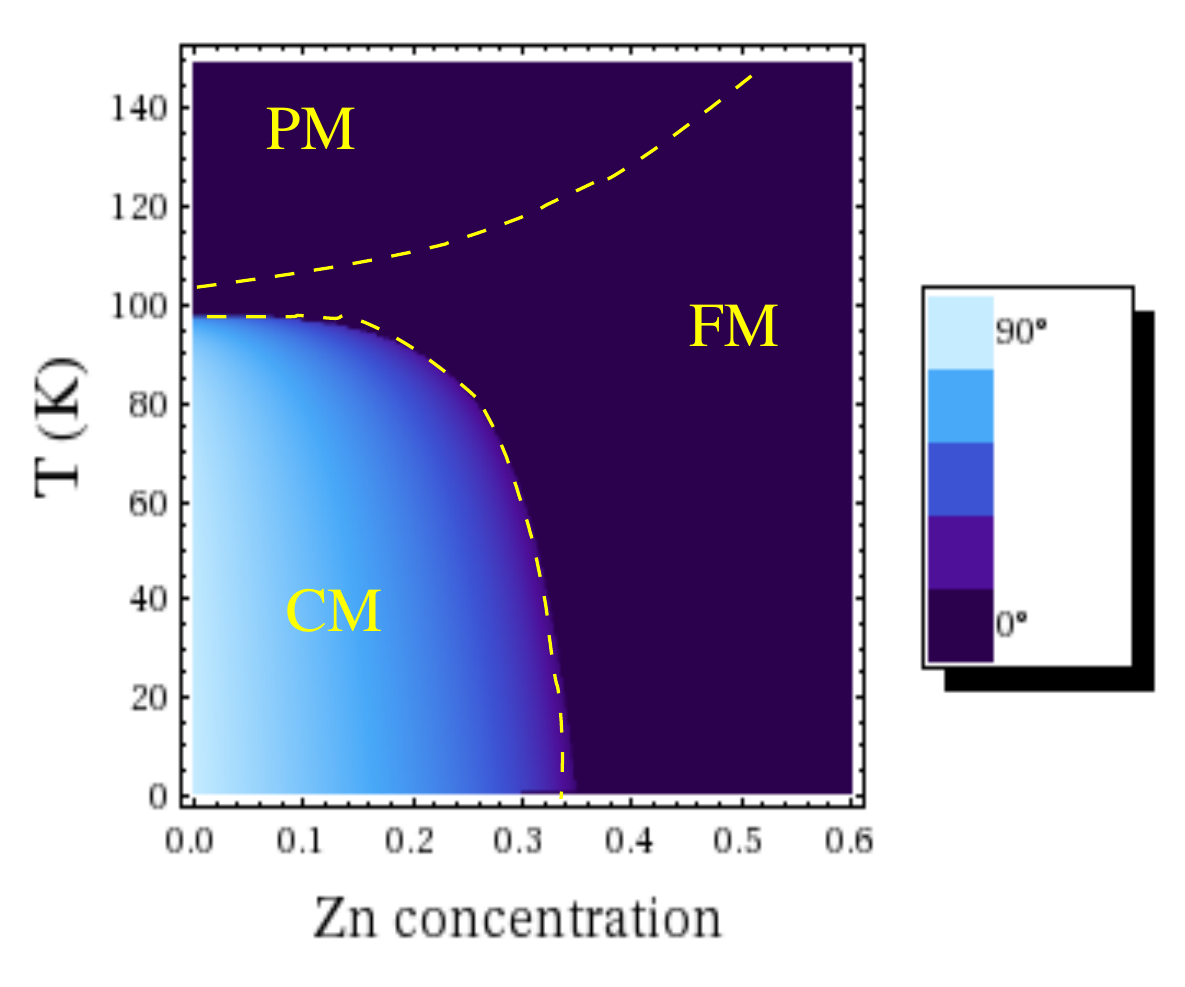}
\caption{(a) The magnetic phase diagram of GdMg, represented by the canting angle $\Theta_c(T)$  and its dependence on 
lattice spacing alongside a schematic picture of the CM state. (b) The magnetic phase diagram of GdMg$_{(1-x)}$Zn$_x$,
 $\Theta_c(T)$, and its dependence on  $x$ for a fixed lattice spacing equal to the 2\% reduction value in (a), equal to $a_{th}$ of  
GdMg$_{0.6}$Zn$_{0.4}$.}
\label{Fig3}
\end{flushleft}
\end{figure}

Fig.2(a) summarises our results. These are the first {\it ab-initio} calculations to show canted 
magnetism (CM) in GdMg. The figure shows the 
emergence of a CM from either a FM ($\Theta_c=0$) 
or AF1 state ($\Theta_c=$180$^{\circ}$). For low $T$, $\Theta_c$
ranges from 70$^{\circ}$ at the theoretical equilibrium lattice constant (0\% reduction)  through to 120$^{\circ}$ (4\% reduction) 
before eventually forming an AF1 magnetic structure 
(angle 180$^{\circ}$) with further reduction.
This agrees very well with experiment~\cite{Liu-GdMg} which finds that, upon lowering the temperature, GdMg orders into a 
FM state at $T_c\approx$ 110K and then undergoes 
a further second order transition into a canted magnetic ordered state at 
$T_F \approx$ 85K~\cite{Liu-GdMg,Morin}. At low $T$ the magnetisation   
$\approx$ 5 $\mu_B$, a value we have also confirmed with our own experimental 
measurements. This is indicative of the FM and AF components,
$m_f$ and $m_a$, being roughly the same size giving a canting angle between  
7 $\mu_B$-sized Gd moments of roughly 90$^{\circ}$.  This state is robust against applied magnetic fields~\cite{Morin} of 
up to 150 kOe. The experimental results are matched almost exactly by our calculations shown in Fig.2(a) for a 2 $\%$ lattice 
spacing reduction from $a_{th}$.
Liu et al.~\cite{Liu-GdMg} also found that under pressure GdMg orders into a AF1 from a PM phase
at $\approx$ 100K  and at a lower temperature undergoes a further first order metamagnetic transition into a canted FM phase. 
The authors estimated the pressure derivative of the magnetisation to be 
-0.04 $\mu_B$ kbar$^{-1}$ at 4.2K, which we have also confirmed experimentally and in fair agreement with
our calculated low $T$ value of -0.02 $\mu_B$ kbar$^{-1}$.

Experimentally it is known that when Gd is replaced by Tb in GdMg, there is a  1\% lattice contraction ~\cite{Aleonard} and a 
FM state undergoes a transition into a canted magnetic structure at low T with $\Theta_c$ 
of at least 90$^{\circ}$. Replacing Gd with 
Dy leads to a larger lanthanide contraction and 
measurements~\cite{Belakhovsky} show that DyMg  orders into 
an AF1 state, developing non-collinear structure with a FM component 
at low $T$ and $\Theta_c$ of 
about 110$^{\circ}$. This correlates with Fig.2(a)~\cite{SI} 
for the smaller lattice spacing regime. The little available data for Ho-Mg~\cite{Aleonard} also indicates canted AF magnetic structure at low $T$. So we infer that the lanthanide contraction~\cite{REnature} 
in part causes the transition from FM-canted to AF-canted magnetic 
structures as the heavy lanthanide series is traversed. 
Our figure 2(a) also implies a tricritical point (PM-AF-FM) at some 
concentration, $y$, in the (Tb$_{1-y}$Dy$_y$)-Mg alloy system with a 
transition to a canted structure at a marginally lower temperature or 
possibly a transition into a canted structure directly.

This unusual canted magnetism of GdMg is evidently destroyed by nominally filled, low-lying 3d or 4d bands  from the non-lanthanide constituent.
 Our calculations, Fig.2(b), show what happens when a fraction $x$ of the Mg sites in GdMg is replaced by Zn. $T_c$ increases with $x$, 
and the low temperature canted structure vanishes altogether for $x > 0.35$. This observation is  
in excellent agreement with the experimental data for GdMg$_{(1-x)}$Zn$_x$ of Buschow et 
al.~\cite{Buschow-Gd-MgZn} who gave an early report of a serious shortcoming of the RKKY picture.

The successful capturing of these unusual temperature and pressure trends of the Gd intermetallics' magnetism is a 
consequence of the theory's detailed description of the valence electrons. The theory includes both the
response of these electrons to the magnetic ordering of the f-electron local moments as well as their effect upon it.
\begin{figure}[t]
\begin{flushleft}
%(a)\includegraphics[width=70mm,height=50mm]{fig_3a.pdf}
%(b)\includegraphics[width=70mm,height=50mm]{fig_3b.pdf}
\includegraphics[trim=20mm 20mm 0 20mm,clip,width=95mm,height=80mm]{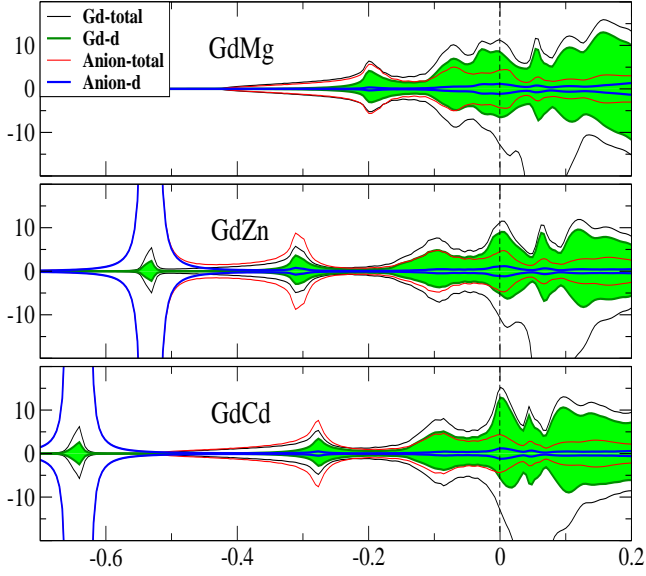}
\caption{The local density of states (DOS) at $a=a_{th}$ for the PM states of  
GdMg, GdZn and GdCd resolved into Gd (black curve) and Mg,Zn or Cd anion (red) components. The Gd d-component
(green curve shaded underneath) and anion d-component  (blue curve) are also
shown. The upper (lower) panel shows the DOS for an electron spin-polarised parallel(anti-parallel) to the local 
moment on the Gd site. The total DOS, an average over all directions, is unpolarised.}
\label{Dos}
\end{flushleft}
\end{figure}
\begin{figure}[t]
\begin{flushleft}
(a)\includegraphics[width=35mm,height=35mm]{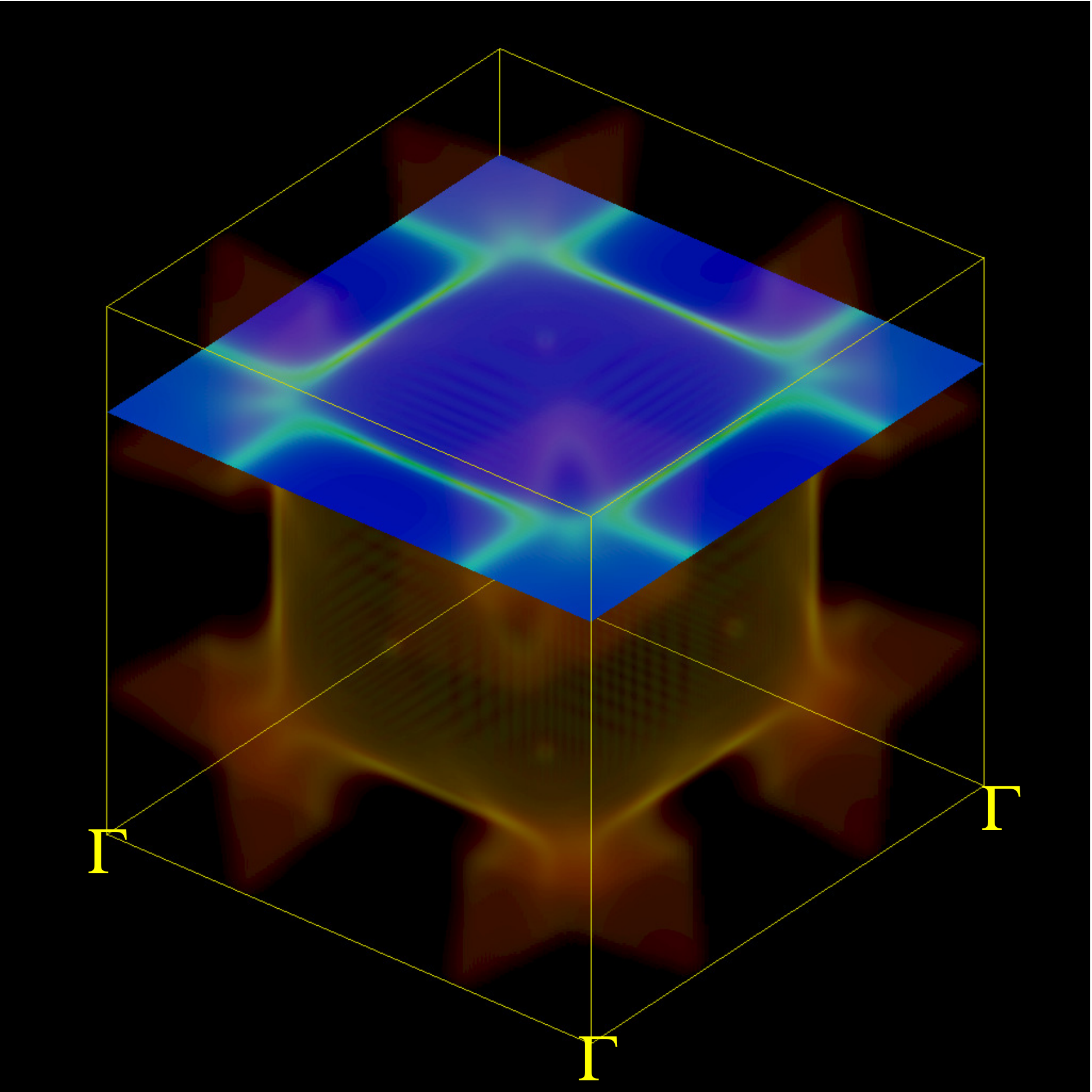} \hspace{0.5cm}
(b)\includegraphics[width=35mm,height=35mm]{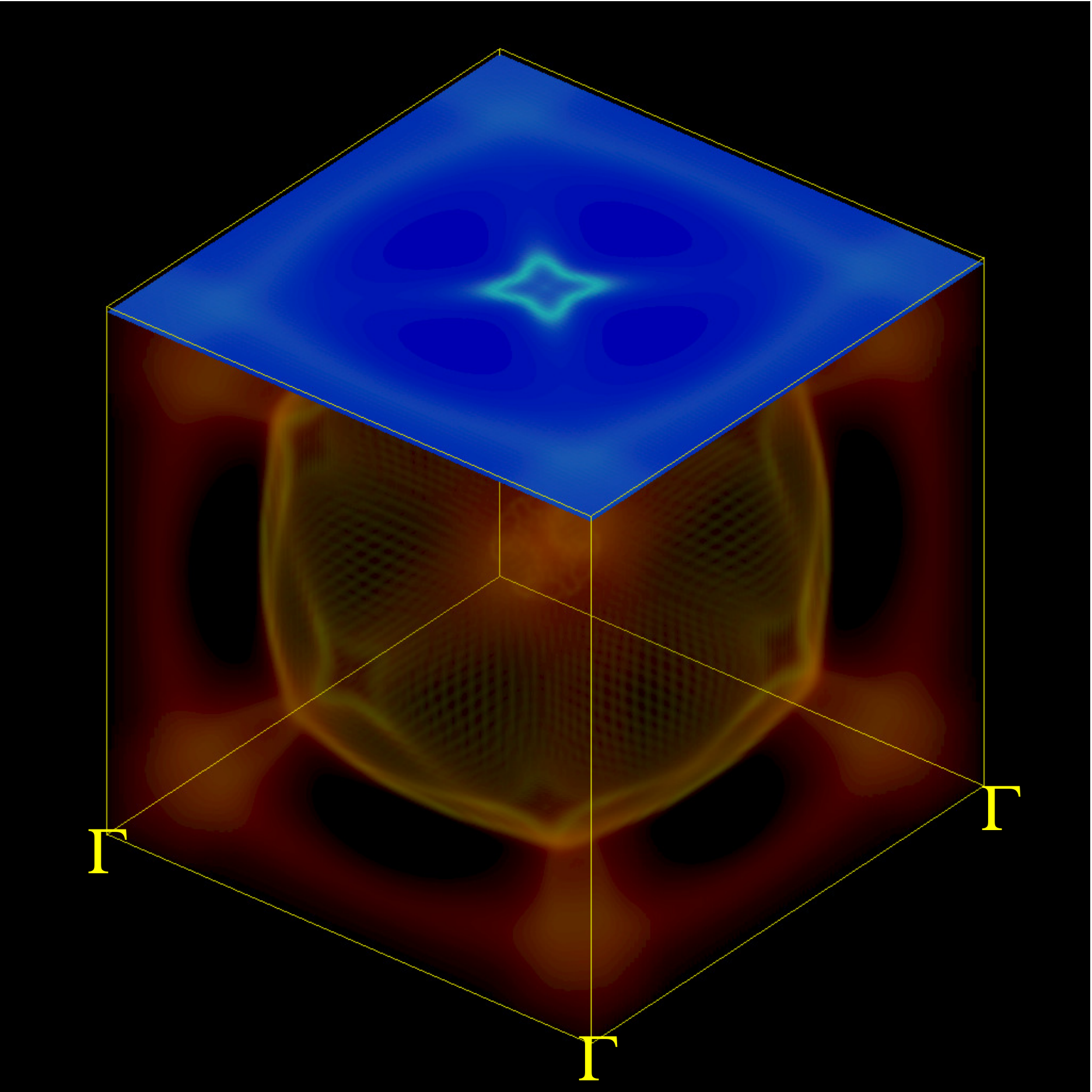}
\caption{The 3D Fermi surface for (a) PM GdMg ($a=0.96a_{th}$, where Fig.2(a) shows AF1 order) which shows nesting, and (b) PM GdCd,
($a=0.94 a_{th}$, close to Fig.1(b)'s peak position) showing the 'hot spot' at wavevectors ${\bf k}= 
(\frac{1}{2}, \frac{1}{2},0)$. The finite width of the FS features reflect the local moment disorder.}
\label{FS}
\end{flushleft}
\end{figure}
Fig.~\ref{Dos} shows the non-free 
electron-like PM valence density of states (DOS) of GdMg, GdZn and GdCd at $a_{th}$ for an electron 
spin-polarised parallel and anti-parallel to the local moment on the Gd site~\cite{Gyorffy_DLM}. Averaged over equally 
weighted moment orientations the DOS is unpolarised overall.  Below $T_c$ the electronic structure adjusts and 
spin-polarises~\cite{SI} when magnetic order develops. The Gd f-moment interactions are 
properties of the electronic structure around the Fermi energy, 
$\varepsilon_{F}$. The Fermi surface (FS) of PM GdMg (for $a=0.96 a_{th}$), Fig.~\ref{FS}(a), shows a distinctive box structure so that a wave-vector, 
$(0,0,\frac{1}{2})$, connects (nests)~\cite{Roth,Dugdale,Dobrich} large portions of parallel FS sheets and drives AF1 
magnetic correlations. This topological feature is absent in GdZn's and GdCd's FS's. Weak hybridization between Gd-5d and lower lying, 
nominally filled Zn-3d or Cd-4d states, shown in Fig.~\ref{Dos}, causes complex differences between their electronic structures around 
$\varepsilon_{F}$ and GdMg's. In GdZn the Zn 3d bands are narrower than GdCd's 4d ones and lie at slightly higher 
energies~\cite{SI}. Moreover we find that lattice compression increases Gd d-state occupation relative to sp-ones in these 
compounds~\cite{Pettifor,Grechnev} which affects FS topology. In particular, as shown in Fig.~\ref{FS}(b) for GdCd, we find 
that Fig.~\ref{pressure}(b)'s peak correlates with a distinct electronic topological 
transition - a 'hot spot' formed by a hole pocket around  ${\bf k}=(\frac{1}{2},\frac{1}{2},0)$, collapsing as $a$ is reduced. 

Atomically localised f-electrons and their intricate physics is 
inevitably the focus for lanthanide material studies. But the valence 
electron glue in which the f-moments sit also harbors surprises. Its 
s-, p- and d-electrons can shift it far from a nearly free electron 
model, as exemplified by the canted magnetism of GdMg and the stark 
contrast of the magnetism of isoelectronic GdZn and GdCd with their 
disparate pressure variations. The predictive {\it ab-initio} computational 
modelling described here has successfully accounted for the subtle 
aspects of the valence electrons' spin polarisability around 
$\varepsilon_F$ and how it is affected by occupation of lower-lying 
lanthanide-other metal d-electron bonding states. This implies that further 
successful quantitative modelling of the rich variety of technologically 
useful lanthanide-transition metal materials must also treat valence 
electronic structure accurately and in quantitative detail. We have shown 
that coordinated {\it ab-initio} theory-experimental studies have the capability 
of producing new guidelines for understanding the magnetism in 
lanthanide-transition metal magnets. Factors such as the average number 
of valence electrons or band-filling, separation in energy of the 
lanthanide 5d and the other constituents' d-bands and the valence band 
widths, reminiscent of the modern analogs of the famous Hume-Rothery 
rules~\cite{alloys} for alloy phase stability, will influence the nature 
of the valence electron structure around $\varepsilon_F$ and the magnetism 
it supports.

{\bf Acknowledgements}\\
We acknowledge the late W.M.Temmerman for valuable and
insightful discussions during the early stages of this work.
The work was supported by the UK EPSRC by Grant No. EP/J06750/1 and an EPSRC service level agreement with 
the Scientific Computing Department of STFC. Work at Ames Laboratory was supported by the Materials Sciences 
and Engineering Division of the Office of Basic Energy 
Sciences of the U.S. Department of Energy under contract no. DE-AC02-07CH11358 with Iowa State University. \\

\end{document}